\documentclass[%
 reprint,
 amsmath,amssymb,
 aps,
]{revtex4-2}

\usepackage{graphicx}
\usepackage[colorlinks,allcolors=blue]{hyperref}
\usepackage{dcolumn}

\begin{document}

\title{Compact linear optical scheme for Bell state generation}

\author{Suren A.\,Fldzhyan}

\author{Mikhail Yu.\,Saygin}%
 \email{saygin@physics.msu.ru}
 
\author{Sergei P.\,Kulik}
 
\affiliation{%
  Quantum Technology Centre and Faculty of Physics, M.V.\,Lomonosov Moscow State University, 119991 Moscow, Russia
}%


\begin{abstract}
    The capability of linear optics to generate entangled states is exploited in photonic quantum information processing, however, it is challenging to obtain entangled logical qubit  states. We report, to the best of our knowledge, the most compact scheme producing the dual-rail-encoded Bell states out of four single photons. Our scheme requires a five-mode interferometer and a single photon detector, while the previously known schemes use six-mode interferometers and two photon detectors. Using computer optimization, we have found a decomposition of the five-mode interferometer with a minimum number of beam-splitters and phase-shift elements. Besides compactness, our scheme also offers a success probability of $1/9$, which is higher than $2/27$ provided by the six-mode counterparts. The analysis suggests that the elevated success probability is connected to higher order of photon interference realized by our scheme, in particular, four-photon interference is implemented in our scheme, while three-photon interference was implemented in previous counterparts.
\end{abstract}

\maketitle

\section*{Introduction}

Linear optics can generate non-classical states of light using initially separable single photons. The simplest example of entangled states generated by linear optics is the one produced by interference of two single photons on a balanced beam-splitter, constituting  the well-known Hong-Ou-Mandel effect~\cite{HOM}. Similarly, large-scale entangled states can also be obtained by linear optics, for which interference of multiple photons in multimode interferometers can be exploited, as in the case of boson samplers~\cite{BosonSampling,PhotonAdvantage}.

However, most commonly performing quantum information tasks requires logical qubit encoding rather just superposition of photons. In turn, this imposes severe constraints on the transformation capabilities of linear optical schemes making  generation of the logical states non-deterministic~\cite{Gubarev2020}, and logical operations over these states even more so~\cite{CarolanScience,DowlingToffoli}. As a result, the generation of high-dimensional entangled multi-qubit states directly in a single multimode linear optical device is impractical.

For this reason, the approaches to generation of high-dimensional multi-qubit states are predominantly based on ways of generating  small-scale resource states that can serve as building blocks of the target states used in quantum communication~\cite{MB_QuantumCommunication,Economou} and quantum computing~\cite{RudolphBallistic,PRX_ResourceCost}. The maximally entangled dual-rail-encoded Bell states, which are the states of our interest in this work, are the easiest and most applicable entanglement resource~\cite{nielsen_chuang_2010}. Another example of a small-scale resource state is the three-photon Greenberger-Horne-Zeilinger~\cite{About_GHZ} state widely used in photonic measurement-based quantum computing~\cite{RudolphBallistic,Psi_fusionbased,Psi_interleaving}.

It is deceptive to think of devising few-qubit photonic experiments as an easy task. In particular, the development of linear optical schemes for generation of two- and three-qubit resource states can be problematic for human intuition due to the extraordinary behaviour of interfering photons, thereby making the development unproductive or suboptimal~\cite{Gubarev2020,Krenn2020}. To aid in designing experiments, currently, quantum information is enjoying the proliferation of computer methods not long ago utilized solely in the classical domain, including methods of machine learning and artificial intelligence~\cite{KrennPRL,Melnikov2018,Erhard2018,Stanisic,Gubarev2020}.

One can categorize the linear optical schemes by access to active feed-forward operations. In particular, for Bell state generation, if active feed-forward is used any generated state differing from the target Bell state by a permutation of modes is considered as a success, since it can be transformed accordingly~\cite{ZhangRudolphPan}. Even though feed-forward yields higher success probability, this comes with errors and losses introduced by active elements into the quantum information algorithm, not to mention the complexity of technical realization. Of course, whether feed-forward should be used or not in a large-scale photonic system is a matter of trade-offs. In contrast, schemes with no feed-forward are designed to generate one particular Bell state~\cite{Stanisic,Gubarev2020}. It is this static type of linear optical schemes, which is of our interest in this work. We use computer optimization to construct the scheme producing dual-rail-encoded Bell states with an aim to find a scheme as compact and efficient as possible.

It is known from previous works that at least four single photons are necessary to generate dual-rail-encoded Bell states by linear optics~\cite{Stanisic,Gubarev2020}. Also, to produce the state a six-mode interferometer and two heralding photon detectors were considered~\cite{Stanisic,Gubarev2020}. We have improved the known schemes in both compactness and success probability, namely, a five-mode scheme requiring one auxiliary mode and one photon detector has been found. 


The paper is organized as follows. In Sec.~\ref{sec:Schemes} we introduce linear optical schemes, which we consider to generate Bell states. In Sec.~\ref{sec:Optimization}, we describe the optimization procedure utilized to find optimal parameters of the schemes. The obtained results are given in Sec.~\ref{sec:Results}. We summarize in Sec.~\ref{sec:Conclusion}.

    \begin{figure*}[htp]
        \centering
        \includegraphics[width=0.95\textwidth]{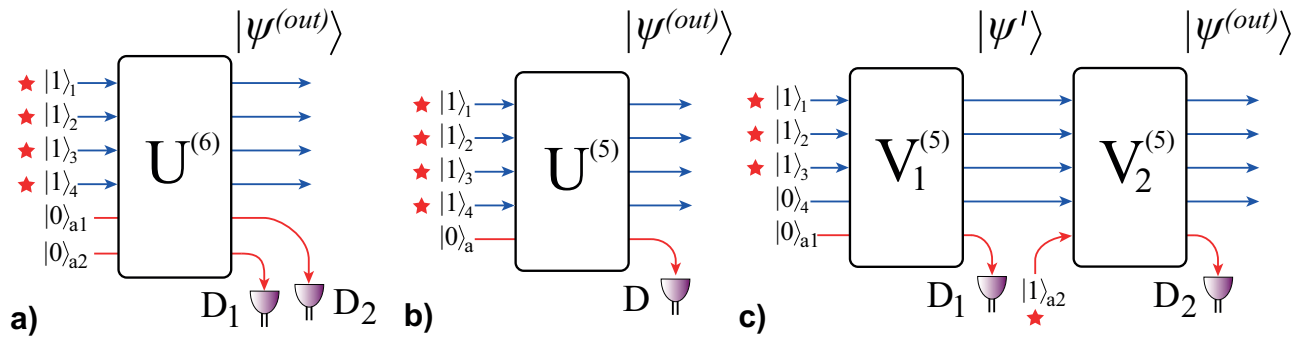}
        \caption{Three types of linear optical schemes for generation of dual-rail-encoded Bell states using four separable photons: a) a scheme that uses a six-mode interferometer $U^{(6)}$ and two heralding photon detectors, D$_1$ and D$_2$. Both heralding detectors should measure single photons for the scheme to generate the Bell state. This type of scheme has been studied previously in~\cite{Gubarev2020}; b) a scheme that uses a five-mode interferometer $U^{(5)}$ and one heralding photon detector D. The heralding detector should measure two photons for the scheme to generate the Bell state; c) a scheme that uses a sequence of two five-mode interferometers and two detectors, D$_1$ and D$_2$. For successful heralding both detectors should measure single photons. }
        \label{fig:fig_1}
    \end{figure*}

\section{Schemes for Bell state generation}\label{sec:Schemes}

There are four maximally entangled Bell states: $|\Psi^{(\pm)}\rangle=(|01\rangle_L\pm|10\rangle_L)/\sqrt{2}$, $|\Phi^{(\pm)}\rangle=(|00\rangle_L\pm|11\rangle_L)/\sqrt{2}$, where $|0\rangle_L$ and $|1\rangle_L$ are the logic basis states of a qubit. States $|\Psi^{(\pm)}\rangle$, $|\Phi^{(\pm)}\rangle$ are connected with each other by local operations so that any of the states can be obtained from another one by applying a single-qubit operator. For example, assuming the state $|\Phi^{(+)}\rangle$ at hand, the other states are derived by: $|\Phi^{(-)}\rangle=\hat{Z}_1|\Phi^{(+)}\rangle$, $|\Psi^{(+)}\rangle=\hat{X}_1|\Phi^{(+)}\rangle$, and $|\Psi^{(-)}\rangle=\hat{X}_1\hat{Z}_1|\Phi^{(+)}\rangle$, where $\hat{X}_j$ and $\hat{Z}_j$ are the Pauli operators with the subscript $j$ referring to a mode the operator is applied to. 

Here, we consider a dual-rail encoding method that requires two spatial modes and one photon per one qubit, so that an arbitrary single-qubit state is a superposition over the physical Fock states: $|0\rangle_L=|1\rangle_1|0\rangle_2$ and $|1\rangle_L=|0\rangle_1|1\rangle_2$, where $|0\rangle_j$ and $|1\rangle_j$ are states with $0$ and $1$ photon in the mode $j$, respectively. The encoding method is conveniently implemented, for example, by an integrated photonics platform, enabling creation of large-scale quantum devices. Also, arbitrary single-qubit operations are implemented easily and deterministically by programmable two-port Mach-Zehnder interferometers. Therefore, dual-rail encoding is the choice in many realizations of quantum information processing tasks~\cite{CarolanScience}. Without loss of generality, as a concrete example we are interested in generation of the Bell state  $|\Phi^{(+)}\rangle$ having the following physical state:
    \begin{equation}\label{eqn:Bell_state}
        |\Phi\rangle=\frac{1}{\sqrt{2}}\left(|1010\rangle+|0101\rangle\right),
    \end{equation}
where the superscript will be omitted in the following.

By contrast, in the dual-rail encoding entangling two-qubit operations are probabilistic and require extra resources of photons and modes of transformations, in addition to those reserved for qubit encoding itself. In particular, it is known that linear optical generation of Bell states requires at least four photons, two of which are to be measured in the auxiliary modes of the optical scheme, wherein a suitable sequence of the detection events will unambiguously herald the target state prepared by the scheme~\cite{Stanisic,Gubarev2020}. We are attempting to improve the Bell states generation process to achieve higher success probability, while using minimal spatial resources. 

Fig.~\ref{fig:fig_1} illustrates three optical schemes considered as means to generate a Bell state. The first scheme depicted in Fig.~\ref{fig:fig_1}a exploits a six-mode linear optical interferometer $U^{(6)}$ that makes four initially separable input photons interfere with each other. It is known from prior works~\cite{Stanisic,Gubarev2020} that the Bell states may be produced in this kind of six-mode schemes when the transfer matrix of the interferometer  is properly chosen and two photon detectors, D$_1$ and D$_2$, are used for heralding. A single photon should be measured at each detector for successful heralding. The probability of a successful detection event was found to be $2/27$. We use this type of scheme as a well-known example to compare the results of other two types of schemes and to assess the correctness of the optimization algorithm described in the following section.

The second scheme, which is depicted in Fig.~\ref{fig:fig_1}b, to the best of our knowledge, has not been considered before. It exploits a five-mode interferometer $U^{(5)}$ and one photon detector D, which makes it more compact than the first one. Since there is an auxiliary mode, two photons can be measured by the detector and removed from the state. Our goal is to prove that an interferometer $U^{(5)}$ which guarantees generation of the Bell state does exist, find its optimal form characterized by as minimal number of elements as possible and highest success probability.

We also introduce the third scheme shown in Fig.~\ref{fig:fig_1}c, in order to study the effect of the order of photon interference on the capability to generate Bell states.  This scheme exploits a sequence of two interferometers, $V_1^{(5)}$ and $V_2^{(5)}$ and photodetection after each of them. Three photons enter the first interferometer that produces an intermediate state  $|\psi'\rangle$, that is partially measured by the detector D$_1$. If one photon is measured, then the procedure proceeds by sending the unmeasured part of the intermediate state to the second interferometer together with one fresh photon. In the end, measuring one photon by the second detector D$_2$ heralds the Bell state at the output of the scheme. By construction, the scheme purposefully limits the number of photons that can interact inside the interferometer to three. Notice that in the first two schemes four photons are allowed to interfere simultaneously.

\section{Optimization procedure}\label{sec:Optimization}

\begin{figure}[htp]
        \centering
        \includegraphics[width=0.4\textwidth]{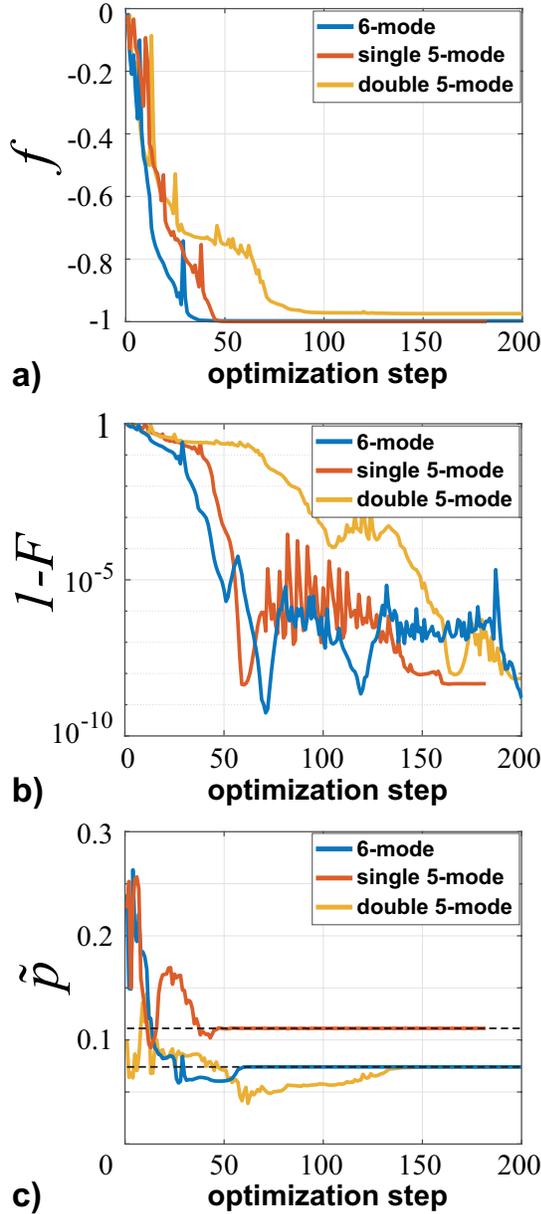}
        \caption{The illustration of the optimization procedure: a) the cost function $f$ dependence on the number of optimization steps, b) the state infidelity $1-F$ and c) probability $\tilde{p}$ in the course of the optimization. The dashed lines mark probability levels of $2/27$ and $1/9$. The following parameters were used in the optimization algorithm:  $\mu=10^{-3}$, $\mu=10^{-4}$, $\mu=10^{-2}$ for the three schemes under study in order they are present in Fig.~\ref{fig:fig_1};  $\varepsilon=10^{-5}$ for all three schemes. }
        \label{fig:fig_2}
    \end{figure}

The state at the output of a lossless $N$-mode interferometer is the superposition over multiple basis states:

    \begin{equation}\label{eqn:state_general_expansion}
        |\psi\rangle_{sa}=\sum_{\mathbf{t}}c_{\mathbf{t},\mathbf{s}}|\mathbf{t}\rangle_{sa},
    \end{equation}
where vectors $\mathbf{s}=(s_1,\ldots,s_N)$ and $\mathbf{t}=(t_1,\ldots,t_N)$ denote the photon occupation vectors at the input and output, respectively. In general, the sum in \eqref{eqn:state_general_expansion} runs through all possible vectors $\mathbf{t}$, the number of which grows as $\begin{pmatrix}
  M+N-1\\ 
  M
\end{pmatrix}$, where $M$ is the number of photons and $N$ is the number of modes. As the total number of photons is conserved by the lossless transformation, it follows that $\sum_js_j=\sum_jt_j=M$. Following the assignment of modes adopted in Fig.~\ref{fig:fig_2}, the basis states are split into a logical (s) and an auxiliary (a) part: $\mathbf{t}=\mathbf{t}_s\oplus\mathbf{t}_a$, so that $|\mathbf{t}\rangle_{sa}=|\mathbf{t}_s\rangle_{s}|\mathbf{t}_a\rangle_{a}$. The probability amplitudes are calculated as  $c_{\mathbf{t},\mathbf{s}}=\text{perm}(U_{\mathbf{t},\mathbf{s}})/\sqrt{\mathbf{t}!\cdot\mathbf{s}!}$, where  $\text{perm}(U_{\mathbf{t},\mathbf{s}})$ is a permanent of the matrix $U_{\mathbf{t},\mathbf{s}}$, which is obtained by taking the columns and rows of the transfer matrix of the interferometer $U$ according to vectors $\mathbf{s}$ and $\mathbf{t}$, respectively \cite{Tichy};   $\mathbf{s}=s_1!\cdot\ldots\cdot{}s_N!$ and $\mathbf{t}=t_1!\cdot\ldots\cdot{}t_N!$ is the shorthand for factorials.

To parametrize the transfer matrices of the interferometers constituent with the schemes depicted in Fig.~\ref{fig:fig_1}, we use the universal decomposition of unitary matrices suggested in~\cite{Clements2016}.  The decomposition represents an $N$-mode interferometer as $N$ layers consisting of beam-splitters with variable transmissivities and variable phase shifts. Each beam-splitter acts locally on two neighboring modes by the transfer matrix:
    \begin{equation}\label{eqn:BSmatrix}
       T_j(\theta,\varphi)=\left(
        \begin{array}{cc}
             e^{i\varphi}\sin\theta & \cos\theta \\
             e^{i\varphi}\cos\theta & -\sin\theta
        \end{array}\right),
    \end{equation}
where $\theta$ specifies the transmissivity of the beam-splitter $\tau=\cos^2\theta$ ($0\le\theta\le\pi/2$), $\varphi$ is the phase difference between the input modes of the beam-splitter ($-\pi\le\varphi<\pi$). The resultant multimode transfer matrix is the product of $Q=N(N-1)/2$ matrices: $U=D\cdot{}T_Q^{(n_Q,m_Q)}\cdot\ldots\cdot{}T_1^{(n_1,m_1)}$, with $D$ being a diagonal matrix of independent phase shifts $\alpha_i$ ($i=\overline{1,N-1}$) and $T^{(m_q,n_q)}_q$ is the beam-splitter matrix acting on modes $m_q$ and $n_q$. It has been proven that this way an arbitrary $N\times{}N$ transfer matrix can be parametrized by setting the phase shifts properly~\cite{Clements2016}.  

Non-deterministic schemes herald the demanded quantum states by a pattern of photons $\mathbf{d}$ detected in the auxiliary modes with a success probability $p$. Thus, it follows that the output quantum state has the form:
    \begin{equation}\label{eqn:optimal_state_form}
        |\Psi_0\rangle_{sa}=\sqrt{p}|\Phi\rangle_s|\mathbf{d}\rangle_a+\sqrt{1-p}|R\rangle_{sa},
    \end{equation}
where $|R\rangle_{sa}$ is the byproduct component of the state always present in non-deterministic linear optical schemes. Of course, for practical reasons, the interferometers should achieve maximal success probability $p$. However, even a small admixture of a component that contradict the dual-rail encoding degrades the quality of the target state. For example, it can introduce errors in the algorithm the state is used for, or make its usage impossible altogether. Hence, the necessary requirement is that the term $|R\rangle_{sa}$ should not contain amplitudes at basis states, whose auxiliary part coincide with $|\mathbf{d}\rangle$, used for successful heralding. 

Also notice that in \eqref{eqn:BSmatrix}, values $\theta=0$ and $\theta=\pi/2$ correspond to the trivial crossover and identity transformations, which are easier to implement in practice than a general two-mode beam-splitter with a specific splitting ratio. Therefore, in addition to the necessary requirements, we also demand that the interferometers are constructed from a minimal number of elements.

To find an optimal decomposition of the interferometer, we use an optimization algorithm that explores the parameter space of the linear optical circuits to find the global minimum of a cost function that account for the aforementioned requirements. Specifically, we used the L-BFGS algorithm, for which a home-made code has been implemented in C++~\cite{BFGS}.

In optimization, the algorithm was exploring the space of phase-shifts that parametrize the interferometer(-s) to minimize the following cost function:
    \begin{equation}\label{eqn:loss_function}
        f(\vec{\theta},\vec{\varphi})=-\tilde{p}^{\mu}F+\varepsilon\sum_{j=1}^{Q}\left(\sin^22\theta_j + \sin^22\varphi_j\right),
    \end{equation}
with $\tilde{p}$ being the probability of obtaining a state with $|\mathbf{d}\rangle_{a}$ in the auxiliary modes calculated as $\tilde{p}=\langle\chi_{\mathbf{d}}|\chi_{\mathbf{d}}\rangle$, where $|\chi_{\mathbf{d}}\rangle=_{a}\langle\mathbf{d}|\Psi^{(out)}\rangle_{sa}$ is the unnormalized state of the logical modes, $F=|\langle\chi_{\mathbf{d}}|\Phi\rangle|^2/\tilde{p}$ is fidelity. Taking the form of the state \eqref{eqn:optimal_state_form} into account, $F$ attains its maximum value of $1$ if and only if $|\chi_{\mathbf{d}}\rangle=|\Phi\rangle$. The second term in \eqref{eqn:loss_function} penalizes the occurrence of non-trivial beam-splitters and phase-shifts in the decomposition. Parameters $\mu$ and $\varepsilon$ control the interplay between the two terms in \eqref{eqn:loss_function} and their values are chosen for better convergence of the algorithm. 


\section{Results}\label{sec:Results}

To prevent the algorithm from sticking to a local minimum and to guarantee convergence of the loss function into the global minimum, it was run several times (typically $\sim3-5$), each time initializing the phase-shifts with random values. Fig.~\ref{fig:fig_2} shows the examples of converging runs used to find the layouts of the interferometers. The dependencies of the cost function \eqref{eqn:loss_function}, which was the optimized function, is shown in Fig.~\ref{fig:fig_2}a. The dependencies of the state infidelity $1-F$ and probability $\tilde{p}$ that accompany the optimized cost function are shown in Fig.~\ref{fig:fig_2}b and Fig.~\ref{fig:fig_2}c, respectively.

\begin{figure}[htp]
        \centering
        \includegraphics[width=0.45\textwidth]{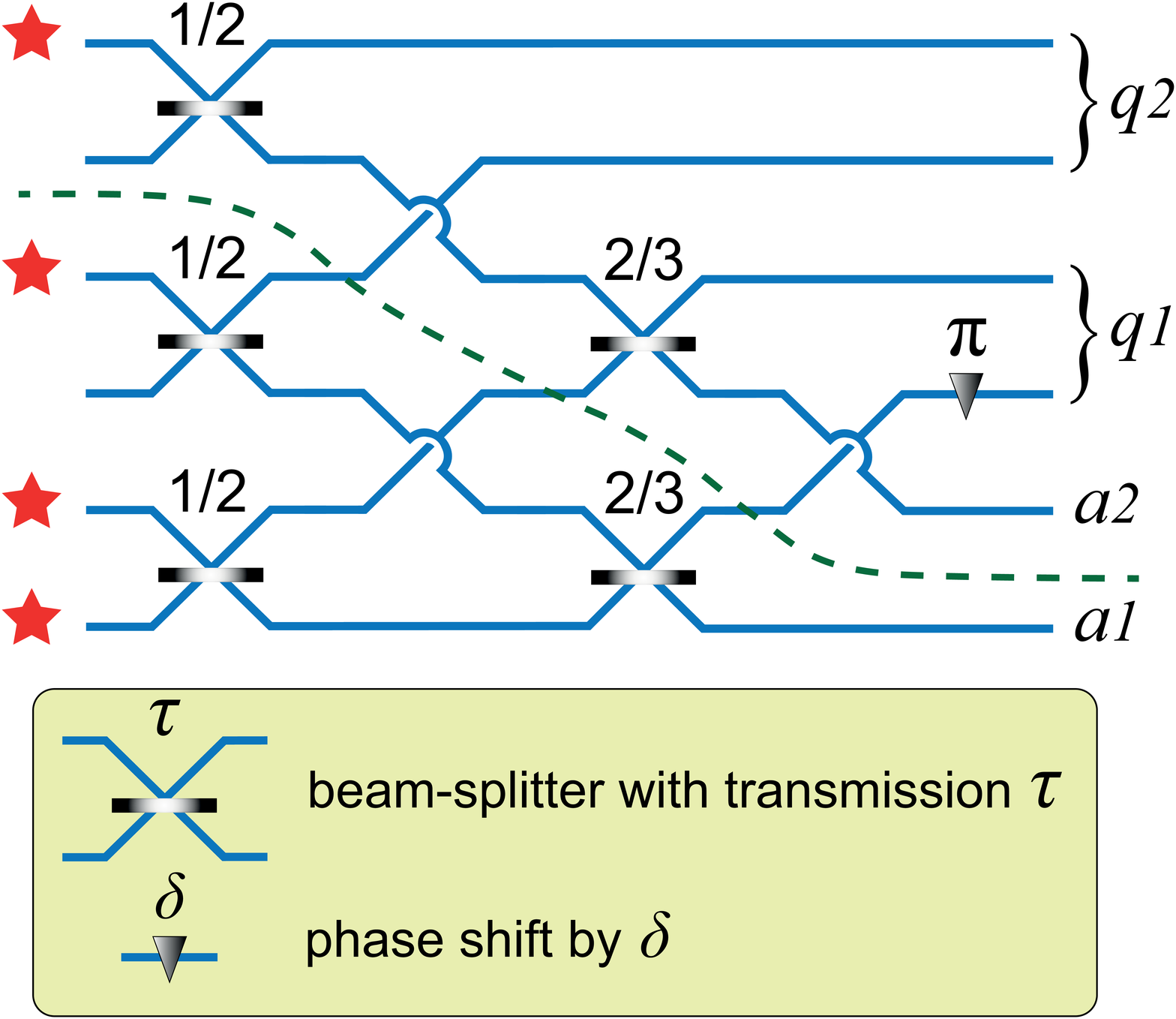}
        \caption{The explicit form of the six-mode scheme obtained by optimization. Here, $q_1$ and $q_2$ are the pairs of modes that encode logical qubits, $a_1$ and $a_2$ are auxiliary modes to be measured by the detectors (see Fig.~\ref{fig:fig_1}a). The beam-splitters are marked by their power transmissivities $\tau_j$. The dashed line separates the scheme into two parts, three-photon interference take place in each of them, provided that one photon is measured in the mode a1. }
        \label{fig:fig_3}
    \end{figure}

First, we found the layout of the six-mode interferometer $U^{(6)}$  (see Fig.~\ref{fig:fig_1}a), shown in Fig.~\ref{fig:fig_3}. It is constructed from two beam-splitters with $\tau=2/3$ and three balanced beam-splitters with $\tau=1/2$. The following state is generated by the interferometer:
    \begin{equation}\label{eqn:psi6}
        |\psi^{(out)}\rangle=\sqrt{\frac{2}{27}}|\Phi\rangle_s|1\rangle_{a1}|1\rangle_{a2}+\frac{5}{\sqrt{27}}|R^{(6)}\rangle_{sa}.
    \end{equation}
Due to cumbersomeness, the explicit form of the residual term $|R^{(6)}\rangle_{sa}$ is given in the Supplementary section (the same is true for the residual terms corresponding to the rest of the schemes). From \eqref{eqn:psi6} it follows that the Bell state $|\Phi\rangle_s$ is generated with probability $p=2/27$ conditioned upon the detection of two single photons by detectors D$_1$ and D$_2$. This result is in agreement with the result of~\cite{Gubarev2020}, however, the difference is in one extra beam-splitter required by the interferometer from~\cite{Gubarev2020} and different placement of the elements. Therefore, we infer that our six-mode scheme is less complex to implement.

Interestingly, the layout of the six-mode interferometer suggests that four photons never interfere simultaneously with each other in case when the heralding detection events take place. To demonstrate this, the dashed line drawn in the layout separates it in two parts. Obviously, one photon entering the upper part is not allowed to get into the lower part. The rest three photons interfere in the lower part and one of them is removed from it in the event of heralding performed on mode a1. This in turn guarantees that a two-photon state goes into the upper part where it interacts with the single-photon state, producing a three-photon state at the output of the logical modes and mode a2.

\begin{figure}[htp]
        \centering
        \includegraphics[width=0.45\textwidth]{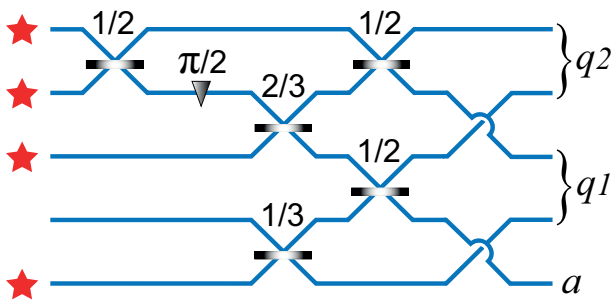}
        \caption{The explicit form of the five-mode scheme obtained by optimization (see Fig.~\ref{fig:fig_1}b). The interferometer provides Bell state generation with probability of $1/9$ heralded by detection of two photons in the auxiliary mode a. 
        The designations are the same as in Fig.~\ref{fig:fig_3}. }
        \label{fig:fig_4}
    \end{figure}

The five-mode interferometer $U^{(5)}$ found by optimization is shown in Fig.~\ref{fig:fig_4}. The output state it produces has the following form:
    \begin{equation}\label{eqn:psi5}
        |\psi^{(out)}\rangle=\frac{1}{3}|\Phi\rangle_s|2\rangle_a+\frac{2\sqrt{2}}{3}|R^{(5,2)}\rangle_{sa},
    \end{equation}
where the first term contains the Bell state \eqref{eqn:Bell_state} heralded by the presence of two photons in the auxiliary mode. The scheme generates the Bell state with probability $1/9$, which is a significant improvement over the previous result. Remarkably, the scheme is built with the same number of beam-splitters and phase-shifts as the six-mode one. It is important to note that, unlike the six-mode scheme, the five-mode scheme exhibits four-photon interference, as one cannot divide the layout into two parts the way we did above. Also, this fact can be proved by the explicit form of the residual term $|R^{(5,2)}\rangle_{sa}$ given by \eqref{eqn:psi5_residual}, which has the probability amplitudes corresponding to all four photons "focused" in one mode.

    \begin{figure*}[htp]
        \centering
        \includegraphics[width=0.95\textwidth]{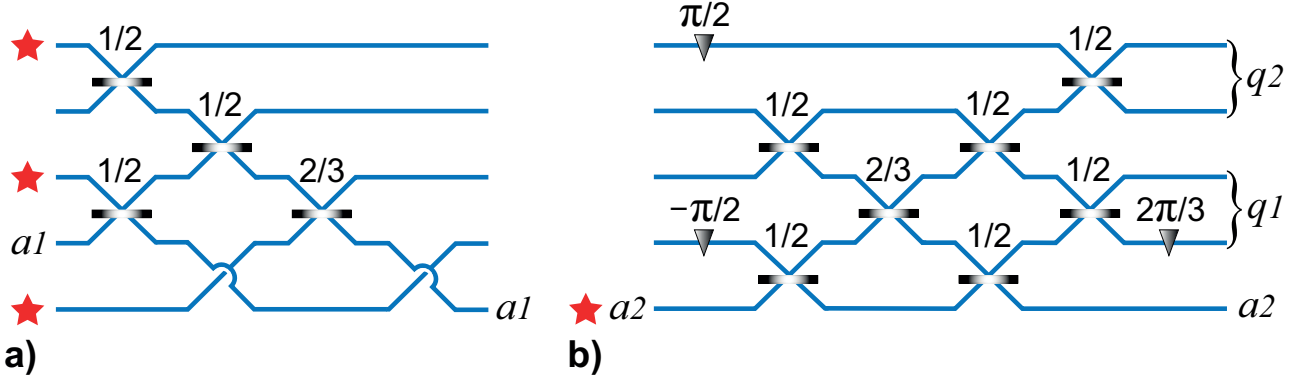}
        \caption{The five-mode interferometers, $V_1^{(5)}$ (a) and $V_2^{(5)}$ (b), forming the scheme depicted in Fig.~\ref{fig:fig_1}c, which were obtained by optimization. The designations are the same as in Fig.~\ref{fig:fig_3}.  }
        \label{fig:fig_5}
    \end{figure*}

The interferometers $V^{(5)}_1$ and $V^{(5)}_2$ of the scheme depicted in Fig.~\ref{fig:fig_1}c  are shown in Fig.~\ref{fig:fig_5}. Three initially separable photons interfere in the first interferometer to produce the intermediate state $|\psi'\rangle$. 
    \begin{equation}
        \begin{split}
        |\psi'\rangle=&\frac{\sqrt{5}}{3}|\chi_0'\rangle_s|0\rangle_{a1}+\frac{1}{3}\sqrt{\frac{5}{2}}|\chi_1'\rangle_s|1\rangle_{a1}+\\
        &+\frac{1}{3}|\chi_2'\rangle_s|2\rangle_{a1}+\frac{1}{3\sqrt{2}}|0000\rangle_s|3\rangle_{a1},
        \end{split}\label{eqn:intermediate_state}
    \end{equation}
where the states $|\chi_j'\rangle_s$ are given explicitly in the Supplemental section (see \eqref{eqn:intermediate_basis}). The auxiliary mode a1 is measured by the photon detector and in the case of a single photon event the generation proceeds by directing the state $|\chi_1\rangle_s$ into the second interferometer together with an extra photon that goes into the auxiliary mode a2. Then the state at the output of the second interferometer has  the following form:
    \begin{equation}
        \begin{split}
            |\psi^{(out)}\rangle=\frac{2}{\sqrt{15}}|\Phi\rangle_s|1\rangle_{a2}+\sqrt{\frac{11}{15}}|R^{(5,1,1)}\rangle_{sa2}.
        \end{split}\label{eqn:psi5_11}
    \end{equation}
As the generation success is conditioned by the single photon detection events at both D$_1$ and D$_2$. It follows from \eqref{eqn:intermediate_state} and \eqref{eqn:psi5_11} that the probability of measuring a single photon by D$_1$ and D$_2$ is $5/18$ and $4/15$, respectively. Accordingly, the  overall probability to generate the Bell state is the product of these values, which yields $2/27$. Therefore, this once again prove that the difference in success probabilities with which linear optical schemes generate the Bell state is connected with the order of photon interference.


\section{Conclusion}\label{sec:Conclusion}

In this work we investigated linear optical generation of the dual-rail-encoded Bell states. Using a numerical optimization algorithm we have improved the six-mode schemes, which were previously considered as the most compact ones, in both compactness and the success probability. Namely, the five-mode scheme that requires a single photodetector has been found. Besides being the most compact known scheme to date, our scheme also offers a much higher generation probability of $1/9$ than in the previous schemes where the generation probability was $2/27$. Along the way, we have shown that the increase in the success probability is connected with the higher order of photon interference implemented by our scheme: four photons interfere in our scheme, while only three photons interfere in the known schemes. These findings highlight the importance of computer methods in designing quantum optical setups. We envision that such methods will be playing a more integral role in the development of practical quantum devices in the future.

\section{Acknowledgments}

S.A.\,Fldzhyan and M.Yu.\,Saygin are grateful to the Foundation for the Advancement of Theoretical Physics and Mathematics (BASIS) for support (Projects No. 20-2-1-97-1 and No. 20-1-3-31-1). The authors acknowledge partial support by the Interdisciplinary Scientific and Educational School of Moscow University ''Photonic and Quantum Technologies. Digital Medicine'' and the  Russian Foundation for Basic Research (RFBR Project No. 19-52-80034).

\section*{Appendix: Residual components of state vectors}

Below we provide the explicit residual terms of the quantum states produces by the schemes, depicted in Fig.~\ref{fig:fig_1} with interferometers circuits shown in Fig.~\ref{fig:fig_3}-\ref{fig:fig_5}.

The residual term of the output state \eqref{eqn:psi6} produced by the six-mode scheme (see Fig.~\ref{fig:fig_3}) has the form:
    \begin{equation}
        \begin{split}
            &|R^{(6)}\rangle_{sa}=\frac{2\sqrt{2}}{5}|\chi_{00}\rangle_s|00\rangle_a+\frac{2}{5}|\chi_{10}\rangle_s|10\rangle_a+\\
            &+\frac{2}{5}|\chi_{01}\rangle_s|01\rangle_a+\frac{\sqrt{2}}{5}|\chi_{20}\rangle_s|20\rangle_a+\\
            &+\frac{\sqrt{2}}{5}|\chi_{02}\rangle_s|02\rangle_a+\frac{1}{5}|\chi_{12}\rangle_s|12\rangle_a+\\
            &+\frac{1}{5}|\chi_{21}\rangle_s|21\rangle_a+\frac{1}{5}|\chi_{30}\rangle_s|30\rangle_a+\frac{1}{5}|\chi_{03}\rangle_s|03\rangle_a+\\
            &+\frac{1}{5\sqrt{2}}|0000\rangle_s|13\rangle_a-\frac{1}{5\sqrt{2}}|0000\rangle_s|31\rangle_a,
        \end{split}\label{eqn:psi6_residual}
    \end{equation}
where the states of the logical modes $|\chi_{ij}\rangle_s$ read
    \begin{equation}
        \begin{split}
            &|\chi_{00}\rangle_s=\frac{\sqrt{3}(|1102\rangle_s-|1120\rangle_s-|1003\rangle_s-|0130\rangle_s)}{4}+\\
            &+\frac{(|1021\rangle_s+|0112\rangle_s+|0031\rangle_s-|0013\rangle_s)}{4},\\
            &|\chi_{01}\rangle_s=\frac{\sqrt{3}|1101\rangle_s+|0111\rangle_s}{2\sqrt{2}}-\frac{|1020\rangle_s+|0030\rangle_s}{2},\\
            &|\chi_{10}\rangle_s=\frac{\sqrt{3}|1110\rangle_s-|1011\rangle_s}{2\sqrt{2}}+\frac{|0102\rangle_s-|0003\rangle_s}{2},\\
            &|\chi_{02}\rangle_s=\frac{\sqrt{3}(|1100\rangle_s+|0011\rangle_s)}{4}+\frac{|0110\rangle_s+3|1001\rangle_s}{4},\\
            &|\chi_{20}\rangle_s=\frac{|1001\rangle_s+3|0110\rangle_s}{4}-\frac{\sqrt{3}(|1100\rangle_s+|0011\rangle_s)}{4},\\
            &|\chi_{12}\rangle_s=\frac{1}{2}|0100\rangle_s+\frac{\sqrt{3}}{2}|0001\rangle_s,\\
            &|\chi_{21}\rangle_s=-\frac{1}{2}|1000\rangle_s+\frac{\sqrt{3}}{2}|0010\rangle_s,\\
            &|\chi_{03}\rangle_s=\frac{1}{2}|0010\rangle_s+\frac{\sqrt{3}}{2}|1000\rangle_s,\\
            &|\chi_{30}\rangle_s=\frac{1}{2}|0001\rangle_s-\frac{\sqrt{3}}{2}|0100\rangle_s.
        \end{split}\label{eqn:psi6_residual_addition}
    \end{equation}

The residual term of the output state \eqref{eqn:psi5} produced by the most compact five-mode scheme (see Fig.~\ref{fig:fig_4}) has the form:

    \begin{equation}
        \begin{split}
            &|R^{(5,2)}\rangle_{sa}=\frac{1}{4}\sqrt{\frac{31}{3}}|\chi_{0}\rangle_s|0\rangle_a+\frac{1}{4}\sqrt{\frac{14}{3}}|\chi_{1}\rangle_s|1\rangle_a+\\
        &+\frac{1}{4}\sqrt{\frac{2}{3}}|\chi_{3}\rangle_s|3\rangle_a+\frac{1}{4\sqrt{3}}|0000\rangle_s|4\rangle_a.
        \end{split}\label{eqn:psi5_residual}
    \end{equation}
where
    \begin{equation}
        \begin{split}
        &|\chi_{0}\rangle_s=\frac{\sqrt{2}}{\sqrt{31}}(|3100\rangle_s-|0130\rangle_s+2|3001\rangle_s-2|0031\rangle_s)+\\
        &+\frac{1}{\sqrt{31}}(|0301\rangle_s-|0400\rangle_s)-\frac{\sqrt{3}}{\sqrt{31}}|1210\rangle_s+\frac{\sqrt{6}}{\sqrt{31}}|1111\rangle_s,\\
        &|\chi_{1}\rangle_s=\frac{1}{\sqrt{7}}(|3000\rangle_s-|0030\rangle_s)+\\
        &+\frac{1}{\sqrt{14}}(|0300\rangle_s-\sqrt{3}|0201\rangle_s)-\sqrt{\frac{3}{7}}|1011\rangle_s,\\
        &|\chi_{3}\rangle_s=\frac{1}{\sqrt{2}}(|0100\rangle_s+|0001\rangle_s).
        \end{split}\label{eqn:psi52_residual_addition}
    \end{equation}

In the scheme with two five-mode interferometers (see Fig.~\ref{fig:fig_5}), the intermediate state $|\psi'\rangle$, produced by the first interferometer $V^{(5)}_1$ reads:
    \begin{equation}
        \begin{split}
        |\psi'\rangle=&\frac{\sqrt{5}}{3}|\chi_0'\rangle_s|0\rangle_{a1}+\frac{1}{3}\sqrt{\frac{5}{2}}|\chi_1'\rangle_s|1\rangle_{a1}+\\
        &+\frac{1}{3}|\chi_2'\rangle_s|2\rangle_{a1}+\frac{1}{3\sqrt{2}}|0000\rangle_s|3\rangle_{a1},
        \end{split}\label{eqn:Appendix_intermediate_state}
    \end{equation}
where $|\chi_j'\rangle$ are the following states:
    \begin{equation}
        \begin{split}
            &|\chi_0'\rangle_s=\frac{1}{2}\sqrt{\frac{3}{5}}(|1110\rangle_s+|0111\rangle_s+|0210\rangle_s)+\\
            +&\frac{1}{\sqrt{10}}(|0021\rangle_s-|1020\rangle_s)+\sqrt{\frac{3}{10}}|1011\rangle_s-\frac{1}{2\sqrt{5}}|0030\rangle,\\
            &|\chi_1'\rangle_s=\frac{1}{2}\sqrt{\frac{3}{5}}(|0101\rangle_s+|1100\rangle_s+|0200\rangle_s+|0020\rangle_s)+\\
            &+\frac{1}{2\sqrt{5}}(|1010\rangle_s-|0011\rangle_s)+\sqrt{\frac{3}{10}}|1001\rangle_s,\\
            &|\chi_2'\rangle_s=\frac{1}{\sqrt{2}}(|0001\rangle_s-|1000\rangle_s).
        \end{split}\label{eqn:intermediate_basis}
    \end{equation}
From \eqref{eqn:Appendix_intermediate_state} it follows that a single-photon event in mode a1, necessary for heralding, is realized with probability $5/18$. Provided that the detector D$_1$ measures a single photon in the mode a1, the state $|\chi_1'\rangle_s$ enters in the second interferometer $V^{(5)}_2$ that finally produces the state \eqref{eqn:psi5_11} with the residual part
    \begin{equation}
        \begin{split}
            &|R^{(5,1,1)}\rangle_{sa}=2\sqrt{\frac{2}{11}}|\chi_0\rangle_s|0\rangle_a+2\sqrt{\frac{2}{11}}|\chi_2\rangle_s|2\rangle_a-\\
            &-\frac{i}{\sqrt{11}}e^{i\alpha}|0000\rangle_s|3\rangle_a
        \end{split}\label{eqn:psi5_11_residual}
    \end{equation}
where
    \begin{equation}
        \begin{split}
            &|\chi_0\rangle_s=\frac{1}{4}\left[|2100\rangle+|0120\rangle-i|0300\rangle-i|0003\rangle\right.+\\
            &-e^{i\alpha}(|2001\rangle-|1002\rangle+|0201\rangle-|0021\rangle+|0012\rangle)+\\
            &+e^{-i\alpha}(|1020\rangle-|2010\rangle-|1200\rangle+|0210\rangle+\\
            &\left.+|0102\rangle-i|3000\rangle+i|0030\rangle)\right],\\
            &|\chi_2\rangle_s=\frac{1}{2}\left[e^{i\alpha}(|0100\rangle+|1000\rangle-|0010\rangle)+e^{-i\alpha}|0001\rangle\right].
        \end{split}
    \end{equation}
and the shorthand $\alpha=\pi/3$ has been introduced to make the expressions more compact.

\bibliography{references}

\end{document}